\documentclass[prl,twocolumn,aps]{revtex4}
\newcommand{\tsg}{\tilde\sigma}
\newcommand{\ts}{i\tilde\omega}
\newcommand{\omt}{\tilde\omega}
\usepackage{epsfig}

\begin{document}
\title{Ac hopping conduction at extreme disorder takes place on the percolating cluster}
\author{Thomas B. Schr{\o}der and Jeppe C. Dyre}
\affiliation{DNRF Centre ``Glass and Time,'' IMFUFA, Department of Sciences, Roskilde University, Postbox 260, DK-4000 Roskilde, Denmark}
\date{\today}

\begin{abstract}
Simulations of the random barrier model show that ac currents at extreme disorder are carried almost entirely by the percolating cluster slightly above threshold; thus  contributions from isolated low-activation-energy clusters are negligible. The effective medium approximation in conjunction with the Alexander-Orbach conjecture leads to an excellent analytical fit to the universal ac conductivity with no nontrivial fitting parameters.
\end{abstract}


\maketitle

Recent advances relating to ion conduction in glasses and other disordered solids include the application of multidimensional NMR techniques \cite{nmr}, the introduction of ac nonlinear spectroscopy \cite{roling}, and elucidations of the high-frequency nearly constant loss \cite{ncl}. Moreover, it was found that the old idea of ions moving by the vacancy mechanism may well be correct \cite{vac}, and simulations gave new insight into the mixed-alkali effect \cite{lam05}. Despite these and other significant advances, important questions remain unanswered. For instance, it is still not understood what role is played by ion interactions for the conductivity \cite{role}, or why the random barrier model (RBM) \cite{rbm,dca} represents ac conductivity data so well. The latter question is not answered below, but new simulations and arguments are presented that we believe lead to a full understanding of the physics of the RBM in the extreme disorder limit (low temperature limit).

Ac conductivity is often studied also for amorphous semiconductors, electronically or ionically conducting polymers, defective crystals of various kinds, polaronic conductors, etc \cite{rbm,dca}. It is a longstanding observation that all disordered solids have remarkably similar ac conductivities \cite{univ}. Universal features include \cite{dca}: At low frequencies the conductivity is constant. At higher frequencies it follows an approximate power law with an exponent less than one that increases slightly with increasing frequency. When measured in a fixed frequency range, the exponent converges to one as temperature goes to zero. The ac conductivity is less temperature dependent than the dc conductivity and obeys time-temperature superposition (sometimes referred to as ``scaling''). The frequency marking onset of ac conduction, $\omega_m$, has the same activation energy as the dc conductivity.

These and other observed features are reproduced by the RBM characterized \cite{dca,dyr} by five assumptions: 1) All charge carrier interactions including self-exclusion are ignored; 2) Charge carrier motion takes place on a cubic lattice; 3) All lattice sites have same energy; 4) Only nearest-neighbor jumps are allowed; 5) Jump rates $\propto\exp(-E/k_BT)$ have random activation energies with distribution $p(E)$. In the RBM the ac conductivity $\sigma(\omega)$ relative to $\sigma(0)$ as a function of a suitably scaled frequency becomes independent of $p(E)$ in the extreme disorder limit, i.e., when the width of $p(E)$ is much larger than $k_BT$ \cite{dca}. Despite lack of non-trivial free parameters the RBM universal ac conductivity gives a good fit to experiment \cite{dca}; more refined models yield results that are close to those of the RBM \cite{rbm_ref}.

It is well-known that the percolation threshold determines the dc conductivity activation energy \cite{perc}. At low temperatures the particles preferably jump across the lowest barriers. The highest barriers on the percolation cluster are bottlenecks dominating the low-temperature dc conductivity. If $E_c$ is the highest barrier on the percolating cluster, one has $\sigma(0)\sim\exp(-E_c/k_BT)$ as $T\rightarrow 0$ \cite{perc}. In order to have a non-zero dc conductivity of the percolation cluster, barriers slightly above the percolation threshold must be included. This defines the ``fat percolation cluster'' \cite{dca}; on length scales shorter than its correlation length the fat percolation cluster appears fractal, on longer length scales it appears homogeneous.

Understanding the RBM universal ac conductivity in terms of percolation arguments is much more challenging. Traditionally \cite{rbm,ac_perc} the problem was approached ``from the high-frequency side'' by proceeding as follows. For (angular) frequencies $\omega\gtrsim\omega_m$ there is a characteristic activation energy $E(\omega)<E_c$ for motion on time scales $\sim 1/\omega$; when $\omega$ decreases towards $\omega_m$ one has $E(\omega)\rightarrow E_c$. Links with $E\le E(\omega)$ form finite low-activation-energy clusters. The cluster size distribution is assumed to determine the ac conductivity. Some time ago we proposed what amounts to coming ``from the low-frequency side,'' namely that all relevant motion takes place on some subset of the infinite percolation cluster \cite{dca}. Numerical evidence for this conjecture is given below, where it is shown that contributions from low-activation energy clusters outside the fat percolating cluster are insignificant. Moreover, it is shown that by assuming that not just a subset, but in fact the entire percolation cluster is important, an excellent analytical approximation to the universal ac conductivity with no nontrivial fitting parameters may be derived. 

The simulations of the RBM reported below refer to the Box distribution of activation energies ($p(E)=1/E_0$ for $0<E<E_0$, zero otherwise); ac universality in the extreme disorder limit implies that this distribution gives representative results \cite{dca}. The lowest temperature simulated is given by $\beta=320$ where $\beta$ is the inverse dimensionless temperature, $\beta\equiv E_0/k_BT$. For $\beta=320$ the jump rates cover more than 130 orders of magnitude, making simulations quite challenging. We used a method based on solving the Laplace transform of the master equation numerically \cite{method}. Conductivity data for $\beta=320$ give an excellent representation of the universal master curve for the RBM over the frequency range studied here \cite{dca}.

\begin{figure}
 \begin{center}
 \includegraphics[width=8.5cm]{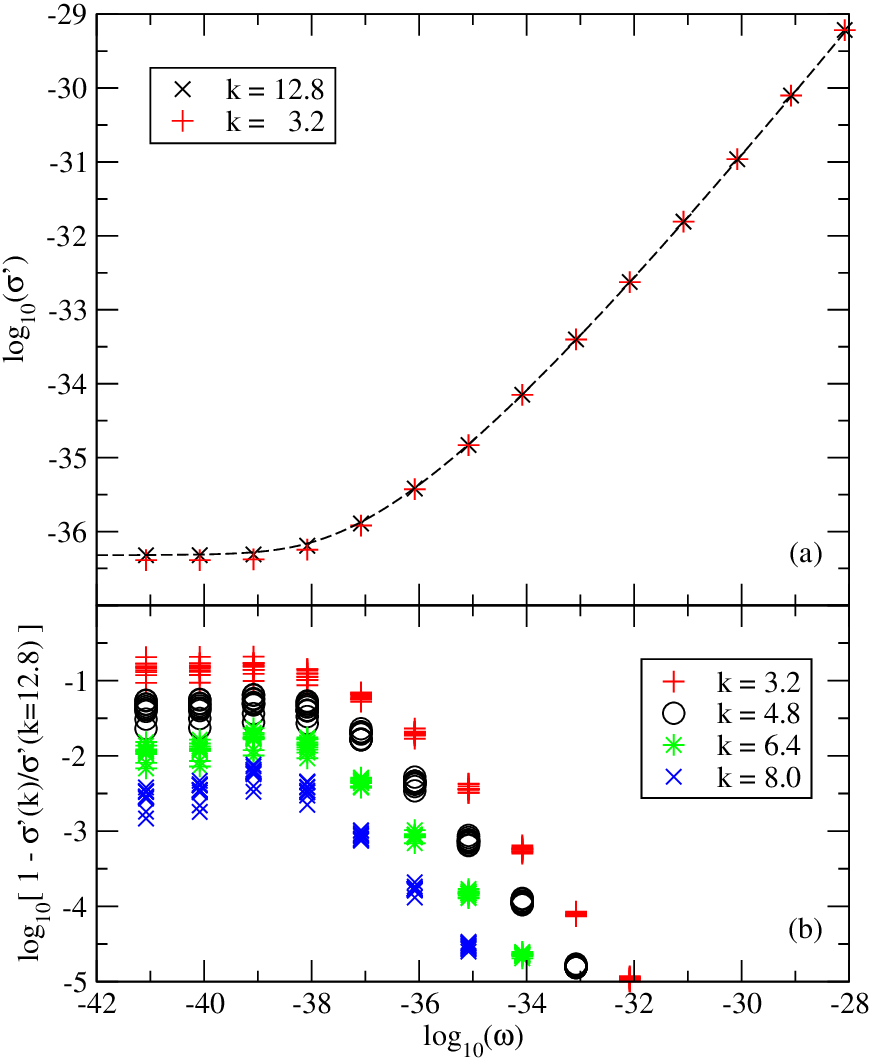}
\end{center}
\caption{Results for the ac conductivity at $\beta\equiv E_0/k_BT = 320$ in rationalized units \cite{dca}; the frequency marking onset of ac conduction, $\omega_m$, is of order $10^{-38}$. Ten independent $96\times 96\times 96$ samples were simulated. (a) Real part of $\sigma(\omega)$ with two  cut-off's: $k=3.2$ and $k=12.8$, averaged over the ten samples. The dashed line is the prediction of Eq. (2) empirically scaled to the $k=12.8$ data. (b) Relative deviation from $k=12.8$ as a function of frequency plotted for each of the ten independent samples.}
\end{figure} 

In previously reported simulations \cite{dca} we applied an activation energy cut-off above the percolation threshold, $E_{\rm cut}/E_0 = E_c+k/\beta$, where $E_c=0.2488$ is the percolation energy for the cubic lattice and $k$ a numerical constant. Jump rates for links with activation energies larger than $E_{\rm cut}$ were set to zero in order to be able to simulate large samples. Figure 1(a) presents the real part of the ac conductivity $\sigma(\omega)=\sigma'(\omega)+i\sigma''(\omega)$ for $k=3.2$ and $k=12.8$ respectively at $\beta=320$. There is little difference between the two data sets. The dashed line gives the prediction of the diffusion cluster approximation (DCA) combined with the Alexander-Orbach conjecture as detailed below (Eq. (\ref{3})). Fig.~1(b) gives the relative errors involved for different $k$ values, taking $k=12.8$ as representing the ``correct'' data. As expected, the errors are largest in the dc regime and decrease with increasing $k$. Choosing $k=6.4$ gives an error of just 1-2\%.

\begin{figure}
 \begin{center}
 \includegraphics[width=8.5cm]{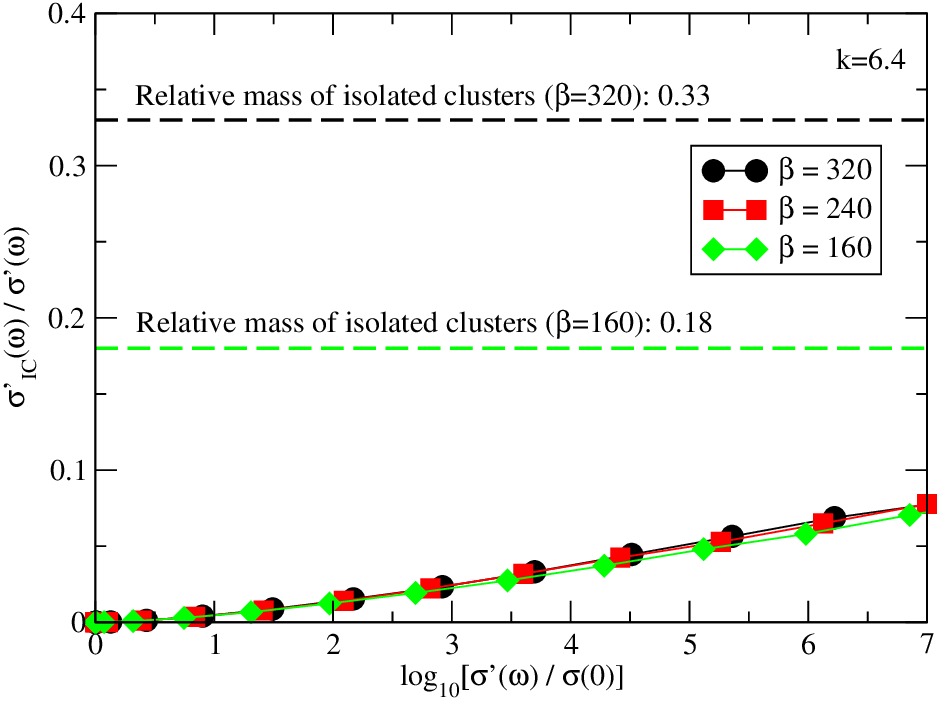}
  \vspace{-.5cm}
\end{center}
\caption{Contribution from isolated clusters for the real part of the ac conductivity, $\sigma'_{IC}(\omega)$, relative to $\sigma'(\omega)$ as a function of the scaled real part of the conductivity, $\tsg'(\omega)\equiv \sigma'(\omega)/\sigma(0)$ (cut-off: $k=6.4$). The two dashed lines mark the relative masses of isolated clusters. Their contribution, however, is much smaller than their relative mass, showing that the dominant part of the ac conduction takes place on the fat percolation cluster.}
\end{figure} 

We proceed to investigate the behavior with $k=6.4$ in more detail. Applying this cut-off,  the links with non-zero jump rate fall into two sets, the ``fat'' percolating cluster and all remaining finite isolated clusters. The latter do not contribute to the dc conductivity. According to the traditional approaches based on cluster statistics, however, they give a significant contribution to the ac conductivity as soon as $\omega\gtrsim\omega_m$ \cite{rbm,ac_perc}. This was never tested numerically. Figure 2 presents the contribution from isolated clusters $\sigma'_{\rm IC}(\omega)$ relative to the full ac conductivity as a function of the real part of the scaled conductivity $\tsg\equiv\sigma(\omega)/\sigma(0)$. The dashed lines mark this relative mass of the isolated clusters for $\beta=160$ and $\beta=320$, respectively. The quantity $\sigma'_{\rm IC}(\omega)/\sigma'(\omega)$, however, is much smaller than the relative mass of isolated clusters for the range of frequencies covered in the figure, i.e., up to 10 billion times $\omega_m$ (compare Fig. 1). Moreover, for $\beta\rightarrow\infty$ the relative mass of isolated clusters goes to one, whereas we find that $\sigma'_{\rm IC}(\omega)/\sigma'(\omega)$ is independent of temperature and stays insignificant. In summary, the dominant part of the low-temperature universal ac conductivity comes from the fat percolation cluster \cite{dca} with little contributions from isolated clusters.

We now turn to the issue of analytical approximations utilizing the effective medium approximation (EMA) \cite{bru35,dyr,note2}. If $G\equiv\int_0^\infty P_0(t)\exp(- i\omega t)dt$ where $P_0(t)$ is the probability for a particle to be at a site given it was there at $t=0$ for a homogeneous system with uniform jump rate, the extreme disorder limit of the EMA self-consistency equation is $\ln\tsg=\Lambda\beta i\omega G$ where $\Lambda$ is a numerical constant \cite{dyr}. This determines a frequency-dependent complex ``effective'' jump rate that is proportional to the frequency-dependent conductivity \cite{rbm,dyr}. Henceforth we switch to the rationalized unit system \cite{dca} where the EMA selfconsistent ac conductivity equals the complex effective jump rate. Because $P_0(t)$ is a function of the effective jump rate times time, $\sigma t$, the quantity $i\omega G$ is a function of $i\omega/\sigma$. In the frequency range relevant for the universal ac conductivity of the extreme disorder limit corresponding to times obeying $\sigma t\gg 1$, one has $|i\omega G|\ll 1$ \cite{dyr,bry80}. If $d$ is dimension, whenever $d\ge 2$ $ i\omega G$ as a function of $i\omega/\sigma$ has a regular first order term \cite{rbm,dyr}: $i\omega G=\alpha_1 (i\omega/\sigma)+...$. If $\Lambda$ is absorbed into a dimensionless frequency by defining $\omt\equiv\alpha_1\Lambda\beta\omega/\sigma(0)$, the EMA universality equation \cite{dyr,bry80} for $d\ge 2$ is

\begin{equation}\label{2}
\tsg\ln\tsg\,=\,
\ts\,.
\end{equation}
This equation gives a qualitatively correct, but numerically inaccurate fit to simulations \cite{dca}. 

In our previous works it was proposed that some unspecified subset of the percolating cluster with fractal dimension $d_f$ (``the diffusion cluster'') is responsible for the ac conduction \cite{dca}. If $d_f<2$ this led to the diffusion cluster approximation (DCA): $\ln\tsg=\left({\ts}/{\tsg}\right)^{d_f/2}$ \cite{dca}. If the diffusion cluster is the so-called backbone, one expects $d_f=1.7$, if the diffusion cluster is the set of red bonds, one expects $d_f=1.1$ \cite{dca}. Treating $d_f$ as a fitting parameter led to $d_f=1.35$ \cite{dca}, however, leaving the nature of the diffusion cluster as an open problem.

What if not just a subset, but in fact the entire percolating cluster contributes significantly to the universal ac conductivity? Random walks on a fractal structure are characterized by $P_0(t)\propto (\sigma t)^{-d_H/2}$ \cite{ben00} where $d_H$ is the spectral dimension. For $d_H<2$ this leads to $i\omega G\propto ( i\omega/\sigma)^{d_H/2}$. In terms of a suitably scaled frequency the EMA thus implies the DCA expression with $d_f=d_H$. According to the Alexander-Orbach  conjecture \cite{ale82} -- known to be almost correct (see, e.g., \cite{hug96,method}) -- one has $d_H=4/3$ for the infinite percolating cluster. If frequency is suitably scaled, this leads to the following approximation to the universal ac conductivity of the extreme disorder limit:

\begin{equation}\label{3}
\ln\tsg\,=\,
\left(\frac{\ts}{\tsg}\right)^{2/3}\,.
\end{equation}

As shown in Fig. 1(a) this expression provides an excellent fit to the universal ac conductivity of the extreme disorder limit \cite{note3}. Equation (\ref{3}) may be put to a more severe test, however, than just fitting the real part of $\tsg(\tilde\omega)$. Figure 3(a) tests one implication of Eq.~(\ref{3}), $|\ln\tsg|=|\omt/\tsg|^{2/3}$, by plotting $|\ln\tsg|$ as function of $|\omt/\tsg|$ in a log-log plot. A cross-over between two power-law regimes is seen, corresponding to a cross-over between Eqs. (\ref{2}) and (\ref{3}). In Fig.~3(b) the apparent exponent $d\ln(|\ln\tsg|)/d\ln|\omt/\tsg|$ is plotted as a function of scaled frequency. Similar results are found by plotting the ratio of the phases of the complex numbers $\ln\tsg$ and $i\omt/\tsg$ (data not shown).

\begin{figure}
 \begin{center}
 \includegraphics[width=8.0cm]{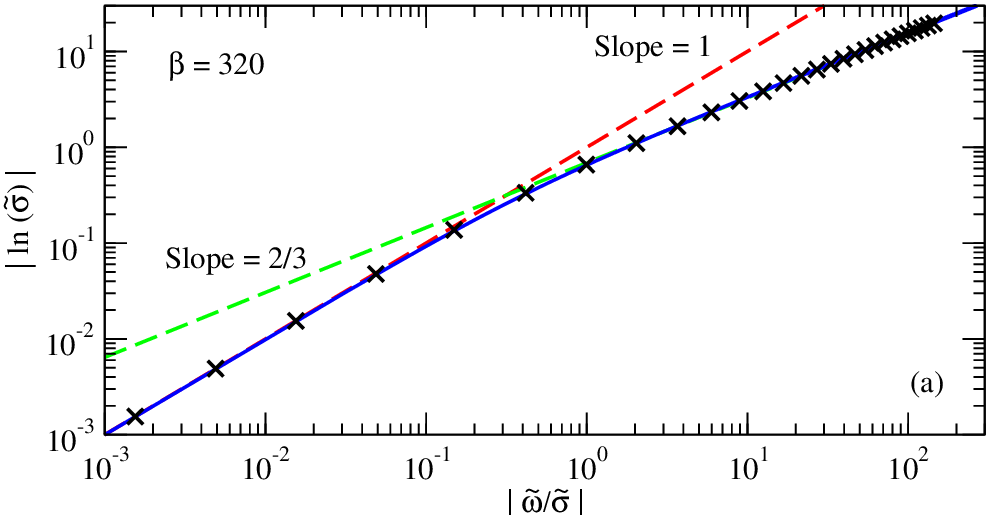}
 \includegraphics[width=8.5cm]{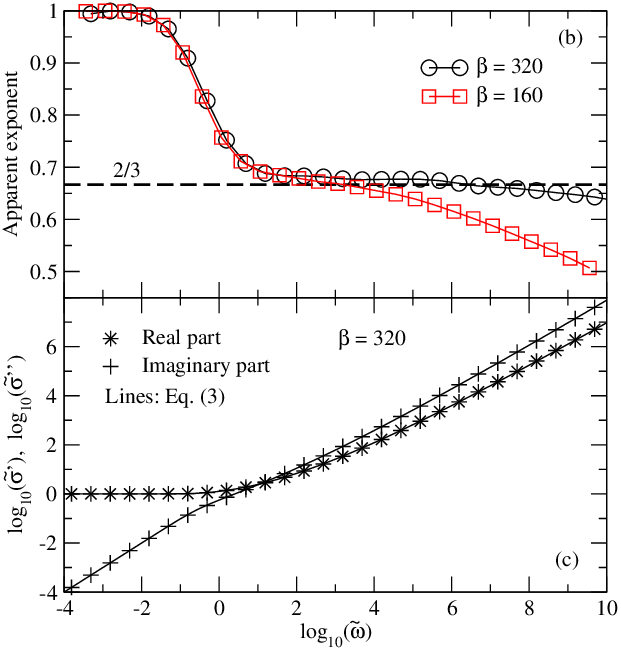}
 \vspace{-.5cm}
\end{center}
\caption{Testing Eqs.~(\ref{3}) and (\ref{4}). Data represent averaging over 100 independent $96\times 96\times 96$ samples ($\beta = 320$, cut-off: $k=6.4$) and 100 independent $64\times 64\times 64$ samples ($\beta = 160$, cut-off: $k=6.4$). For both temperatures the frequency was empirically scaled such that  $\tsg = 1 + \tilde\omega$ in the low frequency limit, where $\tsg\equiv\sigma(\omega)/\sigma(0)$, $\tilde\omega\equiv \omega\Delta\epsilon\epsilon_0/\sigma(0)$ \cite{dca}.
(a) shows $|\ln\tsg|$ as a function of $|\omt/\tsg|$ in a log-log plot. 
(b) shows the apparent exponent $d\ln(|\ln\tsg|)/d\ln|\omt/\tsg|$ as a function of scaled frequency. A cross-over from fractal behavior (exponent $2/3$) to homogeneous behavior (exponent one) is clearly visible. 
(c) shows the real and imaginary parts of the scaled conductivity compared to Eq.(\ref{4}) (full lines).}
\end{figure} 

The picture emerging from Figs.~3(a) and (b) is the following: Equation (\ref{3}) works well whenever $\omt\gtrsim 1$; here the fat percolation cluster appears fractal  because over one cycle the particles move less than the correlation length. At low frequencies there is a transition to the analytic behavior predicted when the dimension is larger than two (Eq. (\ref{2})); over one cycle the particles here move longer than the correlation length and consequently the fat percolation cluster appears homogeneous. 

The entire frequency range is accurately described by the expression

\begin{equation}\label{4}
\ln\tsg\,=\,\frac{\ts}{\tsg} \left( 1 + 2.66 \frac{\ts}{\tsg}\right)^{-1/3} \,
\end{equation}
that is plotted as the full lines in Figs.~3(a) and 3(c). The exponent $-1/3$ was chosen to get agreement with Eq.~(\ref{3}) for $|\omt/\tsg|\gg 1$. The difference between Eqs.~(\ref{3}) and (\ref{4}) is significant only at such low frequencies that  $\tsg(\tilde\omega)$ is of order unity (Fig.~3(a)). Equation~({\ref{3}) breaks down only for the imaginary part for $\omt<1$ where  Eq.~(\ref{3}) predicts $\tsg'' \propto \tilde\omega^{2/3}$ instead of the observed $\tsg '' \propto \tilde\omega$. Numerical solutions of Eqs. (\ref{3}) and (\ref{4}) are provided Ref. \cite{suppl}.

In our opinion, the RBM must now be regarded as solved in the extreme disorder limit in the sense that a good understanding of the model's physics is at hand, leading to an accurate description of the ac conductivity. A notable consequence of the above is that the EMA -- generally believed to be inaccurate except at weak disorder -- works surprisingly well in the extreme disorder limit if the ``geometrical'' input $G$ is taken to reflect the fractal geometry of the percolation cluster. It would be interesting to know whether similar results apply when the EMA is applied for the extreme disorder limit of other models.

\acknowledgments 
The centre for viscous liquid dynamics ``Glass and Time'' is sponsored by the Danish National Research Foundation (DNRF).


\begin{thebibliography}{99}
\bibitem{note2} In Ref. \cite{dyr} we wrote $\tilde G$  for what is here denoted $G$. This reserves the use of tildes to dimensionless quantities.
\bibitem{note3} Equation (\ref{3}) implies that the real part of the ac conductivity a few decades above $\omega_m$ has an approximate power-law frequency depedence with exponents in the range 0.6-0.7, thus close to $2/3$ (compare Fig. 1) and consistent with experiment. It is important to note that this is not related to the exponent $2/3$ appearing in Eq. (\ref{3}); for instance Eq. (\ref{2}) has the same mathematical property.
\end{thebibliography}
\end{document}